\documentclass[12pt]{article}%
\usepackage{amsmath,latexsym}
\usepackage{graphicx}
\usepackage{amsmath}
\usepackage{amsfonts}
\usepackage{amssymb}%
\begin{document}

\title{{Stable wormholes on a noncommutative-geometry
background admitting a one-parameter group of
   conformal motions}}
   \author{
Peter K F Kuhfittig*\\
\footnote{Corresponding Author E-mail: kuhfitti@msoe.edu}
 \small Department of Mathematics, Milwaukee School of
Engineering,\\
\small Milwaukee, Wisconsin 53202-3109, USA}

\date{}
 \maketitle

\begin{abstract}\noindent
When Morris and Thorne first proposed the
possible existence of traversable wormholes,
they adopted the following strategy: maintain
complete control over the geometry, thereby
leaving open the determination of the
stress-energy tensor.  In this paper we
determine this tensor by starting with a
noncommutative-geometry background and
assuming that the static and spherically
symmetric spacetime admits conformal motions.
This had been established in a previous
collaboration with Rahaman et al., using
a slightly different approach.  Accordingly,
the main purpose of this paper is to show
that the wormhole obtained can be made
stable to linearized radial
perturbations. \\

PAC numbers:  04.20.-q, 04.20.Gz
\end{abstract}

\section{Introduction}\label{S:Introduction}

Wormholes are handles or tunnels in spacetime
that connect different regions of our Universe
but may also connect completely different
universes.  As first proposed by Morris and
Thorne \cite{MT88}, wormholes could be actual
physical structures suitable for interstellar
travel.  For the wormhole spacetime they
proposed the following static spherically
symmetric line element:
\begin{equation}\label{E:line1}
ds^{2}=-e^{2\Phi(r)}dt^{2}+\frac{dr^2}{1-b(r)/r}
+r^{2}(d\theta^{2}+\text{sin}^{2}\theta\,
d\phi^{2}),
\end{equation}
using units in which $c=G=1$.  Here
$\Phi=\Phi(r)$ is called the \emph{redshift
function}, which must be everywhere finite
to avoid an event horizon.  The function
$b=b(r)$ is called the \emph{shape function}
since it determines the spatial shape of the
wormhole when viewed, for example, in an
embedding diagram.  The spherical surface
$r=r_0$ is the radius of the \emph{throat}
of the wormhole and must satisfy the
following conditions: $b(r_0)=r_0$, $b(r)<r$
for $r>r_0$, and  $b'(r_0)<1$, usually called
the \emph{flare-out condition}.  This condition
can only be satisfied by violating the null
energy condition, discussed later.

The Einstein field equations in the orthonormal
frame $G_{\hat{\mu}\hat{\nu}}=8\pi T_
{\hat{\mu}\hat{\nu}}$ yield the following simple
interpretation for the components of the
stress-energy tensor:
$T_{\hat{t}\hat{t}}=\rho(r)$, the energy density,
$T_{\hat{r}\hat{r}}=p_r(r)$, the radial pressure,
and $T_{\hat{\theta}\hat{\theta}}=
T_{\hat{\phi}\hat{\phi}}=p_t(r)$, the lateral
pressure.  For the theoretical construction
of the wormhole, Morris and Thorne proposed the
following strategy: retain complete control
over the geometry by specifying the functions
$b=b(r)$ and $\Phi=\Phi(r)$ to obtain the
desired properties and then manufacture or
search the Universe for the materials or fields
that yield the required stress-energy tensor.

Researchers have tried various alternate
strategies to accomplish this goal, two of
which are discussed in this paper.  The first,
the assumption of a noncommutative-geometry
background, is discussed in Sec. \ref{S:non}.
The second, the assumption that the static and
spherically symmetric spacetime admits a
one-parameter group of conformal motions, is
discussed in Sec. \ref{S:Killing}.  Taken
together, these assumptions produce both the
shape and redshift functions and thus a
complete wormhole solution.

Some of these ideas have already been
considered in Ref. \cite{RRKKI}, albeit with
slightly different assumptions and techniques.
What these approaches have in common is a
redshift function resulting in a wormhole that
is not asymptotically flat.  The wormhole
material would therefore have to be cut off
and joined to an external Schwarzschild
vacuum solution.  What makes the redshift
function unusual is that the cut-off can be
chosen in such a way that the junction surface
is a boundary surface, that is, a surface
having zero surface stresses.  The main goal
in this paper is to show that the wormhole
obtained is thereby stable to linearized
radial perturbations.

\section{Noncommutative geometry}\label{S:non}

A particularly useful outcome of string theory
is the recognition that coordinates may become
noncommutative operators on a $D$-brane
\cite{eW96, SW99}.  The commutator is
$[\textbf{x}^{\mu},\textbf{x}^{\nu}]
=i\theta^{\mu\nu}$, where $\theta^{\mu\nu}$ is
an antisymmetric matrix.  The implication is
that noncommutativity replaces point-like
structures by smeared objects.  (The aim is
to eliminate the divergences that normally
occur in general relativity.)  The smearing
effect can be accomplished by using the
Gaussian distribution of minimal length
$\sqrt{\theta}$ instead of the Dirac delta
function \cite{SS03a, SS03b, NSS06, pK13}.
This is also the approach used in Ref.
\cite{RRKKI}.  An equally effective way is
to assume that the energy density of the
static and spherically symmetric and
particle-like gravitational source has
the form \cite{LL12, NM08, KG14, pK15}
\begin{equation}\label{E:rho1}
  \rho(r)=\frac{M\sqrt{\theta}}
     {\pi^2(r^2+\theta)^2}.
\end{equation}
Here the mass $M$ is diffused throughout the
region of linear dimension $\sqrt{\theta}$ due
to the uncertainty.  The noncommutative
geometry is an intrinsic property of spacetime
and does not depend on particular features
such as curvature.

\subsection{Wormhole structure}

The Einstein field equations $G_{\hat{\mu}\hat{\nu}}=
8\pi T_{\hat{\mu}\hat{\nu}}$ result in the
following forms \cite{MT88}:
\begin{equation}\label{E:Ein1}
  \rho(r)=\frac{b'}{8\pi r^2},
\end{equation}
\begin{equation}\label{E:Ein2}
   p_r(r)=\frac{1}{8\pi}\left[-\frac{b}{r^3}+
   2\left(1-\frac{b}{r}\right)\frac{\Phi'}{r}
   \right],
\end{equation}
and
\begin{equation}\label{E:Ein3}
   p_t(r)=\frac{1}{8\pi}\left(1-\frac{b}{r}\right)
   \left[\Phi''-\frac{b'r-b}{2r(r-b)}\Phi'
   +(\Phi')^2+\frac{\Phi'}{r}-
   \frac{b'r-b}{2r^2(r-b)}\right].
\end{equation}
Since Eq. (\ref{E:Ein3}) can be obtained
from the conservation of the stress-energy tensor,
$T^{\mu\nu}_{\phantom{\mu\nu};\nu}=0$, only two
of Eqs. (\ref{E:Ein1})-(\ref{E:Ein3})
are independent.  As a result, these can be
written in the following form:
\begin{equation}\label{E:E1}
  b'=8\pi\rho r^2,
\end{equation}
and
\begin{equation}\label{E:E2}
  \Phi'=\frac{8\pi p_rr^3+b}{2r(r-b)}.
\end{equation}
Eq. (\ref{E:rho1}) immediately yields
the total mass-energy $M$ of the wormhole
inside a sphere of radius $r$:
\begin{equation}\label{E:mass}
   M=\int^r_{0}\rho(r')4\pi (r')^2
   dr'=\frac{2M}{\pi}\left(\text{tan}^{-1}
   \frac{r}{\sqrt{\theta}}-
   \frac{r\sqrt{\theta}}{r^2+\theta}\right).
\end{equation}
Next, from Eq. (\ref{E:Ein1}),
\begin{equation}\label{E:shape1}
   b(r)=\frac{8M\sqrt{\theta}}{\pi}
   \int^r_{r_0}\frac{(r')^2dr'}{[(r')^2+\theta]^2}
   +r_0,
\end{equation}
so that $b(r_0)=r_0$, as required.  Since Eq.
(\ref{E:shape1}) is valid for any $r_0$, the
wormhole can be macroscopic.  The resulting
shape function is
\begin{equation}\label{E:shape2}
  b(r)=\frac{4M\sqrt{\theta}}{\pi}
  \left(\frac{1}{\sqrt{\theta}}\text{tan}^{-1}
  \frac{r}{\sqrt{\theta}}-\frac{r}{r^2+\theta}-
  \frac{1}{\sqrt{\theta}}\text{tan}^{-1}
  \frac{r_0}{\sqrt{\theta}}+\frac{r_0}{r_0^2
  +\theta}\right)+r_0.
\end{equation}
Moreover, from
\begin{equation}\label{E:bprime}
  b'(r)=\frac{8M\sqrt{\theta}r^2}
  {\pi (r^2+\theta)^2},
\end{equation}
we find that (since $\sqrt{\theta}\ll M$)
\[
     0<b'(r)<1,
\]
so that the flare-out condition is met.

\emph{Remark:}  For later use, we observe
that $b(r)$ has continuous derivatives of
all orders.

\subsection{The null energy condition}

Closely related to the flare-out condition is
the violation of the null energy condition.
Observe that
\begin{multline}
  \rho(r)+p_r(r)=\frac{M\sqrt{\theta}}
  {\pi^2(r^2+\theta)^2}+\left.\frac{1}{8\pi}
  \left[-\frac{b}{r^3}+2\left(1-\frac{b}{r}
  \right)\frac{\Phi'}{r}\right]\right|
  _{r=r_0}\\
  =\frac{M\sqrt{\theta}}
  {\pi^2(r_0^2+\theta)^2}-\frac{1}{8\pi}
  \frac{b(r_0)}{r_0^3}<0
\end{multline}
since $\sqrt{\theta}\ll M$.  The violation
of the null energy condition can be attributed
to the noncommutative-geometry background
without hypothesizing the need for ``exotic
matter."

\section{Conformal Killing vectors}
    \label{S:Killing}
We assume in this paper that our static spherically
symmetric spacetime admits a one-parameter group of
conformal motions.  These are motions along which
the metric tensor of a spacetime remains invariant
up to a scale factor.  This assumption is equivalent
to the existence of conformal  Killing vectors such
that
\begin{equation}\label{E:Lie}
   \mathcal{L_{\xi}}g_{\mu\nu}=g_{\eta\nu}\,\xi^{\eta}
   _{\phantom{A};\mu}+g_{\mu\eta}\,\xi^{\eta}_{\phantom{A};
   \nu}=\psi(r)\,g_{\mu\nu},
\end{equation}
where the left-hand side is the Lie derivative of the
metric tensor and $\psi(r)$ is the conformal factor.
(For further discussion, see \cite{MM96, BHL07}.)
The vector $\xi$ generates the conformal
symmetry and the metric tensor $g_{\mu\nu}$ is
conformally mapped into itself along $\xi$.
According  to \cite{HPa, HPb}, this type of symmetry
has been used effectively to describe relativistic
stellar-type objects, not only yielding new solutions,
but leading to new geometric and kinematical insights
\cite{MS93, Ray08, fR10, fR12b}.

As already noted, exact solutions of traversable
wormholes admitting conformal motions are discussed
on Ref. \cite{RRKKI}.  Two earlier studies assumed
\emph{non-static} conformal symmetry
\cite{BHL07, BHL08}.

To study the effect of conformal symmetry, it is
convenient to use an alternate form of the metric
\cite{RRKKI}:
\begin{equation}\label{E:line2}
   ds^2=- e^{\nu(r)} dt^2+e^{\lambda(r)} dr^2
   +r^2( d\theta^2+\text{sin}^2\theta\, d\phi^2).
\end{equation}
The Einstein field equations then take on the
following form:

\begin{equation}\label{E:Einstein1}
e^{-\lambda}
\left(\frac{\lambda^\prime}{r} - \frac{1}{r^2}
\right)+\frac{1}{r^2}= 8\pi \rho,
\end{equation}

\begin{equation}\label{E:Einstein2}
e^{-\lambda}
\left(\frac{1}{r^2}+\frac{\nu^\prime}{r}\right)-\frac{1}{r^2}=
8\pi p_r,
\end{equation}

\noindent and

\begin{equation}\label{E:Einstein3}
\frac{1}{2} e^{-\lambda} \left[\frac{1}{2}(\nu^\prime)^2+
\nu^{\prime\prime} -\frac{1}{2}\lambda^\prime\nu^\prime +
\frac{1}{r}({\nu^\prime- \lambda^\prime})\right] =8\pi p_t.
\end{equation}

Next, following Herrera and Ponce de Le\'{o}n
\cite{HPa}, we restrict the vector field by requiring
that $\xi^{\alpha}U_{\alpha}=0$, where $U_{\alpha}$
is the four-velocity of the perfect fluid
distribution.  The assumption of spherical symmetry then
implies that $\xi^0=\xi^2=\xi^3=0$ \cite{HPa}.  Eq. (\ref
{E:Lie}) now yields the following results:
\begin{equation}\label{E:sol1}
    \xi^1 \nu^\prime =\psi,
\end{equation}
\begin{equation}\label{E:sol2}
   \xi^1  = \frac{\psi r}{2},
\end{equation}
and
\begin{equation}\label{E:sol3}
  \xi^1 \lambda ^\prime+2\,\xi^1 _{\phantom{1},1}=\psi.
\end{equation}

We can then use these equations to obtain
\begin{equation} \label{E:gtt}
   e^\nu  =C r^2
\end{equation}
and
\begin{equation}\label{E:grr}
   e^\lambda  = \left(\frac {a} {\psi}\right)^2,
\end{equation}
where $C$ and $a$ are integration constants.  The
Einstein field equations can also be rewritten:
\begin{equation}\label{E:E1}
\frac{1}{r^2}\left(1 - \frac{\psi^2}{a^2}
\right)-\frac{2\psi\psi^\prime}{a^2r}= 8\pi \rho,
\end{equation}
\begin{equation}\label{E:E2}
\frac{1}{r^2}\left( \frac{3\psi^2}{a^2}-1
\right)= 8\pi p_r,
\end{equation}
and
\begin{equation}\label{E:E3}
\frac{\psi^2}{a^2r^2}
+\frac{2\psi\psi^\prime}{a^2r} =8\pi p_t.
\end{equation}
As before, only two of Eqs. (\ref{E:E1})
-(\ref{E:E3}) are independent.

It is interesting to note that the shape function,
Eq. (\ref{E:shape2}), can also be obtained by
substituting Eq. (\ref{E:rho1}) into Eq.
(\ref{E:E1}) instead of Eq. (\ref{E:Ein1}), as
before.  So the assumption of conformal symmetry
is clearly not needed for determining $b(r)$, but
it does determine the redshift function
from  Eq. (\ref{E:gtt}), thereby completing the
wormhole solution.  Using the notation in Sec.
\ref{S:Introduction}, we therefore have
\begin{equation}\label{E:Phi}
   e^{2\Phi}=Cr^2\quad \text{and} \quad
   \Phi=\frac{1}{2}\,\text{ln}\,(Cr^2).
\end{equation}
It now becomes apparent that our wormhole
spacetime is not asymptotically flat.  In fact,
the redshift function obtained above does not
even appear to have a particularly desirable
form.  It turns out, however, that the form
actually has some unexpected and potentially
useful properties, as we will see in the next
section.

\section{Junction to an external vacuum solution}

Since by Eq. (\ref{E:gtt}) the wormhole spacetime
is not asymptotically flat, the wormhole material
must be cut off at some $r=a$ and joined to an
exterior Schwarzschild solution,
\begin{equation}
ds^{2}=-\left(1-\frac{2M}{r}\right)dt^{2}
+\frac{dr^2}{1-2M/r}
+r^{2}(d\theta^{2}+\text{sin}^{2}\theta\,
d\phi^{2}).
\end{equation}
Here $M=\frac{1}{2}b(a)$, so that
$e^{2\Phi}=Ca^2=1-2M/a$, whence
\begin{equation}
   C=\frac{1-2M/a}{a^2},
\end{equation}
where $M$ is determined from in Eq. (\ref{E:mass}).

While the metric is now continuous at the
junction surface, the derivatives are not.  The
following forms, proposed by Lobo \cite{fL04,
fL05a}, are suitable for present purposes and
will be discussed further in the next section:
\begin{equation}\label{E:sigma1}
   \sigma =-\frac{1}{4\pi a}\left(\sqrt
   {1-\frac{2M}{a}}-\sqrt{1-\frac{b(a)}{a}}
   \right)
\end{equation}
and
\begin{equation}\label{E:P1}
   \mathcal{P}=\frac{1}{8\pi a}\left(
   \frac{1-\frac{M}{a}}{\sqrt{1-\frac{2M}{a}}}
   -[1+a\Phi'(a)]\sqrt{1-\frac{b(a)}{a}}\right).
\end{equation}
Since $b(a)=2M$, the surface stress-energy
$\sigma$ is zero.  From $e^{2\Phi}=Cr^2$,
we find that $\Phi'(a)=1/a$ and $a\Phi'(a)=1$.
Now Eq. (\ref{E:P1}) becomes
\begin{equation}\label{E:P2}
   \mathcal{P}=\frac{1}{8\pi a}\left(
    \frac{1-\frac{M}{a}}{\sqrt{1-\frac{2M}{a}}}
   -2\sqrt{1-\frac{b(a)}{a}}\right).
\end{equation}
Since $b(r)<r$ and $b(r)/r$ is monotone
decreasing, we can choose $a$ in such a way
that $b(a)/a=2/3$, and since $2M=b(a)$, we
get $\mathcal{P}=0$ by Eq. (\ref{E:P2}).

We conclude that the cut-off can be chosen
in such a way that the surface stresses are
zero.  Such a surface is called a \emph{
boundary surface} and should result in a
stable structure.  That is the topic of
the next section.

\section{Stability analysis}

Given the junction surface $r=a$, denoted by $S$,
our starting point is the Darmois-Israel formalism \cite{wI66, mV95}:
if $K_{ij}$ is the extrinsic curvature across $S$ (also known as
the second fundamental form), then the stress-energy tensor
$S^i_{\phantom{i}j}$ is given by the Lanczos equations:
\begin{equation}\label{E:Lanczos}
  S^i_{\phantom{i}j}=-\frac{1}{8\pi}\left([K^i_{\phantom{i}j}]
   -\delta^i_{\phantom{i}j}[K]\right),
\end{equation}
where $[X]=\lim_{r\to a+}X-\lim_{r\to a-}X=X^{+}-X^{-}.$  So
$[K_{ij}]=K^+_{ij}-K^-_{ij},$ which expresses the discontinuity in the
second fundamental form, and $[K]$ is the trace of
$[K^i_{\phantom{i}j}]$.

In terms of the energy-density $\sigma$ and the surface pressure
$\mathcal{P}$, $S^i_{\phantom{i}j}=\text{diag}(-\sigma, \mathcal{P},
 \mathcal{P}).$  The Lanczos equations now yield
\begin{equation}\label{E:stress1}
  \sigma=-\frac{1}{4\pi}[K^\theta_{\phantom{\theta}\theta}]
\end{equation}
and
\begin{equation}\label{E:stress2}
  \mathcal{P}=\frac{1}{8\pi}\left([K^\tau_{\phantom{\tau}\tau}]
    +[K^\theta_{\phantom{\theta}\theta}]\right).
\end{equation}

A dynamic analysis can be obtained by letting the radius $r=a$ be a
function of time, as in Ref.~\cite{PV95}. Here the overdots denote
derivatives with respect to $\tau$.  According to Lobo \cite{fL05b},
the components of the extrinsic curvature are given by
\begin{equation}\label{E:exterior1}
  K^{\tau+}_{\phantom{\tau}\tau}=\frac{\frac{M}{a^2}+\overset{..}{a}}
  {\sqrt{1-\frac{2M}{a}+\overset{.}{a}^2}},
\end{equation}
\begin{equation}\label{E:exterior2}
  K^{\tau-}_{\phantom{\tau}\tau}=\frac{\Phi'\left(1-\frac{b(a)}{a}
     +\overset{.}{a}^2\right)+\overset{..}{a}
    -\frac{\overset{.}{a}^2\left[b(a)-ab'(a)\right]}{2a[a-b(a)]}}
        {\sqrt{1-\frac{b(a)}{a}+\overset{.}{a}^2}},
\end{equation}
and
\begin{equation}\label{E:exterior3}
  K^{\theta+}_{\phantom{\theta}\theta}=\frac{1}{a}
    {\sqrt{1-\frac{2M}{a}+\overset{.}{a}^2}},
\end{equation}
\begin{equation}\label{E:exterior4}
  K^{\theta-}_{\phantom{\theta}\theta}=\frac{1}{a}
    {\sqrt{1-\frac{b(a)}{a}+\overset{.}{a}^2}}.
\end{equation}
These forms yield
\begin{equation}\label{E:sigma}
\sigma=-\frac{1}{4\pi}(K^{\theta+}_{\phantom{\theta}\theta}-
       K^{\theta-}_{\phantom{\theta}\theta})=
   -\frac{1}{4\pi a}\left(\sqrt{1-\frac{2M}{a}+\overset{.}{a}^2}-
   \sqrt{1-\frac{b(a)}{a}+\overset{.}{a}^2}\right).
\end{equation}
In the static case, that is, when $\overset{.}{a}=
\overset{..}{a}=0$, Eq. (\ref{E:sigma}) reduces to Eq.
(\ref{E:P1}).

Again following Lobo \cite{fL05b}, rewriting Eq.~(\ref{E:sigma}) in
the form
\begin{equation}\label{E:modified}
  \sqrt{1-\frac{2M}{a}+\overset{.}{a}^2}=
      \sqrt{1-\frac{b(a)}{a}+\overset{.}{a}^2}-4\pi\sigma a
\end{equation}
will yield the following equation of motion:
\begin{equation}\label{E:motion}
   \overset{.}{a}^2+V(a)=0.
\end{equation}
Here $V(a)$ is the potential, which can be put into the following
convenient form:
\begin{equation}\label{E:potential1}
  V(a)=1-\frac{\frac{1}{2}b(a)+M}{a}
   -\left(\frac{M-\frac{1}{2}b(a)}{m_s}\right)^2
   -\frac{m^2_s}{4a^2}
\end{equation}
where $m_s=4\pi a^2\sigma$ is the mass of the junction surface,
normally referred to as a \emph{thin shell} \cite{fL05b}.  This
applies only to cases in which $\sigma$ is different from zero.

When linearized around a static solution at $a=a_0$, the solution is
stable if, and only if, $V(a)$ has a local minimum value of zero at
$a=a_0$, that is, $V(a_0)=0$ and $V'(a_0)=0$, and its graph is
concave up: $V''(a_0)>0$.  According to Ref. \cite{LC05}, for
$V(a)$ in Eq.~(\ref{E:potential1}), these conditions are met.

Since we are now dealing with a dynamic analysis, we will take
$a=a_0$ to be the cut-off discussed earlier.  Returning to Eqs.
(\ref{E:sigma1}) and (\ref{E:P1}), $\sigma(a_0)=0$ and
$\mathcal{P}(a_0)=0$.  Next, consider the ``boundary layer"
extending from $r=a_0$ to $r=a_0+\epsilon$, where $\epsilon >0$
is an arbitrary constant.  So for $a_1$ in the interval
$(a_0,a_0+\epsilon)$, $\sigma(a_1)$ is no longer equal to zero.
Being arbitrarily thin, the boundary layer can be viewed as a
thin shell separating the interior and exterior solutions.
Since the thin shell is not part of the interior solution,
we have, as a result,
\[
   M<\frac{1}{2}b(a_1)\quad \text{whenever}
            \quad \sigma(a_1)<0
\]
and
\[
    M>\frac{1}{2}b(a_1) \quad \text{whenever}
            \quad\sigma(a_1)>0.
\]

We are seeking a solution that (1) meets the stability
criterion and (2) is arbitrarily close to our solution.
This nearby solution should be viewed as a mathematical
solution, rather than a physical one.

To construct such a solution, we modify $b=b(r)$ on the
interval $(a_0,a_0+\epsilon)$ smoothly, that is, we
modify $b(r)$ in such a way that the second derivative
remains continuous (referring to the earlier Remark).
The modification is shown in Fig. 1: the concave down
\begin{figure}[tbp]
\begin{center}
\includegraphics[width=0.8\textwidth]{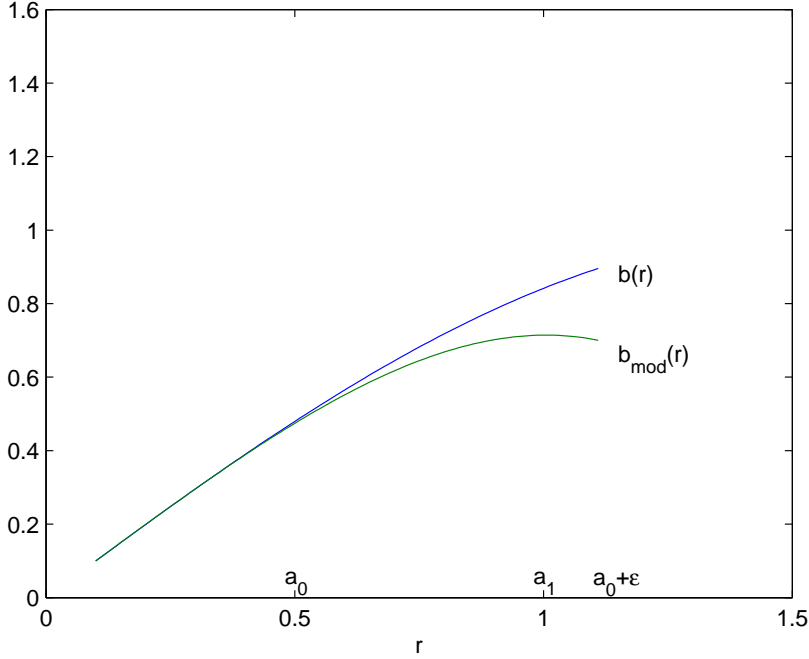}
\end{center}
\caption{$b''(a_1)<0$ for $a_1$ in the interval
      $(a_0,a_0+\epsilon)$.}
\end{figure}
curve $b(r)$ is bent down slightly to produce
$b'_{mod}(a_1)=0$ for some $a_1$ in the interval
$(a_0,a_0+\epsilon)$, where $b_{mod}$ denotes the
modified shape function.  Observe that $b_{mod}$
is also concave down, but it has a larger
curvature, that is, $|b''(r)|<|b''_{mod}(r)|$.
Regarding the notation, modifying $b(r)$ will
also modify $V(r)$ on the interval
$(a_0,a_0+\epsilon)$, denoted by
$V_{mod}(r)$.  Since our goal is to show that
$V''(a_0)>0$ and since $V_{mod}(a_0)=V(a_0)$
and $V_{mod}''(a_0)=V''_{mod}(a_0)$, we will
retain the original notations $b(r)$ and
$V(r)$ to simplify the discussion.

It is important to note that since $\epsilon >0$
can be arbitrarily small, $a_1$ is arbitrarily
close to $a_0$.  Moreover, since $\sigma$ is
nonzero, $m_s=4\pi a_1^2\sigma$ is also nonzero
and, above all, approximately constant.  This
allows us to find $V'$ and $V''$ from Eq.
(\ref{E:potential1}):
\begin{equation}
   V'(a_1)=\frac{d}{da_1}\left(
   -\frac{\frac{1}{2}b(a_1)+M}{a_1}\right)
   -\frac{2}{m_s^2}\left(M-\frac{1}{2}b(a_1)\right)
   \frac{d}{da_1}\left(M-\frac{1}{2}b(a_1)\right)
   +\frac{1}{2}\frac{m_s^2}{a^3}
\end{equation}
and
\begin{multline}
  V''(a_1)=\frac{d^2}{da_1^2}\left(
  -\frac{\frac{1}{2}b(a_1)+M}{a_1}\right)\\
  -\frac{2}{m_s^2}\left\{\left(M-\frac{1}{2}b(a_1)\right)
  \frac{d^2}{da_1^2}\left(M-\frac{1}{2}b(a_1)\right)
  +\left[\frac{d}{da_1}\left(M-\frac{1}{2}b(a_1)\right)
  \right]^2\right\}-\frac{3}{2}\frac{m_s^2}{a^4}\\
  =-\frac{1}{2}\frac{1}{a_1}b''(a_1)
  +\frac{1}{a_1^2}b'(a_1)-\frac{1}{a_1^3}b(a_1)
  -\frac{1}{a_1^3}(2M)\\
  -\frac{2}{m_s^2}\left\{-\frac{1}{2}b''(a_1)
  \left(M-\frac{1}{2}b(a_1)\right)
   +\left[-\frac{1}{2}b'(a_1)\right]^2\right\}
   -\frac{3}{2}\frac{m_s^2}{a^4}.
\end{multline}
We would like to determine the conditions for
which $V''(a_0)>0$.  To that end, we first find the
conditions for which
\begin{equation}\label{E:potential2}
    V''(a_1)>0.
\end{equation}
First observe that $b'(a_1)=0$, $2M\approx b(a_1)$,
while $-(3/2)m_s^2/a^4$ is negligible.  This leads
to
\begin{equation}
    \frac{1}{2}\frac{1}{a_1}b''(a_1)
    <-\frac{1}{a_1^3}(2b(a_1))+\frac{1}{m_s^2}
    b''(a_1)\left(M-\frac{1}{2}b(a_1)\right)
\end{equation}
or, after multiplying both sides by $2a_1$ and
rearranging,
\begin{equation}\label{E:master}
   \frac{2a_1}{m_s^2}\left(\frac{1}{2}b(a_1)-M\right)
   b''(a_1)+b''(a_1)<-\frac{4}{a_1^2}b(a_1).
\end{equation}
To satisfy this inequality, the left side must be
negative.  This implies that we must have
$M<\frac{1}{2}b(a_1)$, so that $\sigma <0$.  The
need for a negative surface density is consistent
with earlier studies involving thin shells \cite
{pK12, aU10, RK10}.)  For the critical left-most
term, we have from $m_s=4\pi a_1^2\sigma$,
\begin{equation}
   \frac{2a_1(\frac{1}{2}b(a_1)-M)b''(a_1)}
   {[4\pi a_1^2]^2\left(-\frac{1}{4\pi a_1}\right)^2
   \left(\sqrt{1-\frac{2M}{a_1}}-\sqrt{1-\frac{b(a_1)}{a_1}}\right)
    \left(\sqrt{1-\frac{2M}{a_1}}-\sqrt{1-\frac{b(a_1)}{a_1}}\right)},
\end{equation}
 which simplifies to
 \begin{equation}
    \frac{b''(a_1) \left(\sqrt{1-\frac{2M}{a_1}}
    +\sqrt{1-\frac{b(a_1)}{a_1}}\right)}
    {\sqrt{1-\frac{2M}{a_1}}-\sqrt{1-\frac{b(a_1)}{a_1}}}.
 \end{equation}
 This (negative) term becomes arbitrarily large in absolute
 value as $a_1\rightarrow a_0$.  So reversing the steps,
 we conclude that sufficiently close to $a=a_0$, inequality
 (\ref{E:master}) and hence inequality (\ref{E:potential2})
 are satisfied.

 Returning to Eq. (\ref{E:potential1}), observe that since
 $M$ and $m_s$ are constants, $V''$ is continuous in its
 domain.  Moreover, since $a_1$ is arbitrarily close to $a_0$,
 $V''(a_1)>0$ implies that $V''(a_0)>0$: for every $a_1$ in
 the interval $(a_0,a_0+\epsilon)$, there exists $\epsilon_1>0$
 such that $a_1=a_0+\epsilon_1$.  So $V''(a_0+\epsilon_1)>0$
 implies that $V''(a_0)>0$ because
 $\lim_{\epsilon_1\to 0}V''(a_0+\epsilon_1)=V''(a_0)$
 by the definition of continuity.

 We conclude that our wormhole is stable to linearized
 radial perturbations.

\section{Conclusions}

The theoretical construction of Morris-Thorne
wormholes is based on the following strategy: retain
complete control over the geometry by specifying
the redshift and shape functions and then manufacture
or search the Universe for materials or fields that
produce the required stress-energy tensor.  This
paper addresses this problem by assuming a
noncommutative-geometry background, thereby
producing the shape function.  The assumption that
the spacetime admits a one-parameter group of
conformal motions then yields the redshift
function.  Adding the assumption of the
conservation of mass-energy completes the
determination of the stress-energy tensor.
From a physical standpoint, then, the redshift
and shape functions have been determined
from the given conditions, rather than simply
assigned.  The result is a complete 
wormhole solution.

The resulting wormhole spacetime is not
asymptotically flat and has to be cut off
and joined to an external vacuum solution
at some junction interface.  The  unusual
nature of the redshift function has yielded
an unexpected conclusion: the cut-off can
be chosen in such a way that the resulting
junction surface is a boundary surface,
having zero surface stresses.  It is
subsequently shown that the resulting
wormhole has another important physical
property: it can be made stable to
linearized radial perturbations.

\end{document}